# Towards a Generic Application Partitioning and Retraction Framework for Pervasive Environments

Nevin Vunka Jungum and Nawaz Mohamudally
School of Innovative Technologies and Engineering
University of Technology Mauritius (UTM)
La Tour Koenig, Mauritius

Nimal Nissanke
School of Computing, Information Systems and Mathematics
London South Bank University, London, UK

*Abstract*— Current mobile context-aware applications for pervasive environments have been designed to consume information from computational nodes or devices in their surroundings or environments. As the hardware industry continues making much smaller, compact and cheap hardware, the vision of having plenty of very small powerful digital networking nodes in, for e.g., the living room or bedroom, is not so far. Designing software that can make optimal use of all these computational nodes when needed is still challenging; since software will not only consume information from these nodes but parts of the software can be hosted on these different nodes. In this paper we propose the *BubbleCodes Framework* which is a generic application partitioning and retraction framework for next generation context-aware applications that will have the capabilities to partition and retract themselves on multiple computational nodes in a pervasive environment.

## I. INTRODUCTION

The scientific community witnessed an increase in interest in the area of context-awareness since the publication of the work of *Roy Want et al.* [1] in 1992 that utilized "location" information to improve their "Active Badge Location System" project. However, the motivation behind might have come from one of the most cited paper in computer science, "The Computer for the 21$^{st}$ Century" by *Mark Weiser* [2] (5358 citations as reported by Google Scholar [3]) that enable researchers to think about a new way of interacting with computers. The famous words of "Ubiquitous Computing" later also popularly known as "Pervasive Computing" was coined.

Context-Aware Computing or Context-Awareness is a fundamental core ingredient for the building up of ubiquitous or pervasive computing systems or environments. The picture was made clearer in a paper entitled "Context-Aware Computing Applications" [4] published in 1994, where the authors emphasize at the start itself that context-aware computing applications describe software that examines and reacts to an individual's changing context. Thereafter, there have been several attempts at defining context-awareness [5][6] and context [7][8]. It is today widely acknowledged that applications that are context-aware considers information like location, time, identity, activity, user preferences, temperature, among many others, to better adapt the system to the user or in other words to enable the user to focus in his/her task instead of manipulating the technologies involved. Many projects [9][10][11] have taken advantage of the different sensors or computational nodes available in a smart or pervasive environment. Once information is collected and modeled by an application's context aggregator [12], the application can thus enhance its adaptability and awareness of the current situation.

Talking about mobile applications running in pervasive environments, one way of viewing things is to think about applications running in mobile devices, for e.g., smart-phones, getting information from sensors embedded in the environment – context information, and finally adapt the application's behavior accordingly. This is where most of the current works are centered at. We argue that in a pervasive environment, next-generation context-aware applications need not only retrieve data from sensors or computational nodes/devices but be able to make use of the nodes/devices' resources as well. In other words we envision next-generation context-aware applications to have the capabilities to decompose themselves into multiple blocks of codes, move to the nodes/devices in the environment to make use of their resources and come back to their originating location as and when required. Research in this direction recently started (mid 2008), where the authors [13] proposed a prototype application for communication in a pervasive environment called *"ContextCom"*. The authors reported that the application has the capabilities of breaking itself into different codes fragments (partitioning property), move to other devices in the environment and come back to the originating device (retraction property).

Partitioning an application into multiple *code-blocks* and deciding in which node/device to host which *code-block* is a complex task if there is no protocol set. Now imagine of the complexity when dealing with a real-life pervasive environment, for e.g., a smart house, where there are multiple users and multiple applications running concurrently. Adding to it is the hot spice of "mobility" which is an undeniable factor in a pervasive environment. Since users are highly expected to be mobile, so network disconnection is no longer "an assumption" but "a core fundamental principle" that needs to be taken into consideration. Taking into consideration the above discussed issues, we propose in this paper the definition and design of a generic application partitioning and retraction framework that would ensure that context-aware applications will have the partitioning and retraction properties. In Section II we discuss existing works in this direction and Section III describe a motivating scenario showing the usefulness of partitioning and retraction. The framework is covered in Section IV followed by a discussion in Section V. In Section VI we conclude with some comments on our immediate future works.

## II. RELATED WORKS

Research in software partitioning has been ongoing since 1997 starting with the JavaParty [15] project. Since then there has been numerous works, for e.g., Doorastha [16], Addistant [17], J-Orchestra [18], HYDRA [19] and so on. However, all of these works are meant for fixed networks, mobility is not taken into consideration, and most importantly are not resilient to network disconnection.

As mentioned in the introduction, the idea of partitioning and retraction of applications for pervasive environments was investigated in a recent paper [13] in the mid 2008. In this work [13][14] the authors proposed the idea of resilient actor model whereby the functionality of an application is decomposed in a set of resilient actors which are modular and active program units. In this approach, partitioning refers to moving the actors to the different available nodes (fixed/mobile) in the environment.

The work is an interesting first step towards partitioning and retracting applications in pervasive environments. However, we believe that instead of jumping directly into how to do the partitioning process and what resilient strategy to choose, etc, we think it is more appropriate to understand the overall software, user and device interaction and to build up a framework. Since apart from network resilience, we noticed several important issues not addressed like handling of context information by the partitioned blocks of codes, multi-applications environment, multi-user environment, communication between partitioned blocks of codes, devices' privileges and user privacy among many others.

## III. MOTIVATING SCENARIO AND ANALYSIS

To demonstrate the usefulness of partitioning and retraction, we describe the running of a video player in a pervasive environment.

Figure 1 depicts four situations. Three rooms are shown with their respective resources, that is, in the bedroom with have a desktop computer, in the kitchen we have a HiFi system, and in the living room we have a HiFi system and television.

In the first situation, 1 (a), a user is watching a movie on his/her smartphone. As he moves to his bedroom, second situation 1(b), the application (video player) detects a resource of the user, that is, his desktop computer. Depending on the user preference settings, the video player decomposes itself and moves to the desktop computer while leaving a component on the smartphone. Thus the user can make maximum use of the resources (monitor to view, speakers to output sound and keyboard for movie controlling purposes) of the desktop computer without draining out the battery of his smartphone.

In the third situation, 1(c), the user decides to go to the kitchen. Once the user starts walking away from the bedroom along with his smartphone and out of the network range of the desktop computer, the application (video player) immediately retract back and continue to play on the smartphone without the intervention of the user.

Once in the kitchen, the video player detects a new resource, a HiFi system that has the capability to output sound. It then decomposes itself and moves part of the application responsible for outputting the sound to the HiFi system to make use of the speakers of the HiFi system. However, the application's parts for viewing the movie and controlling it (forward, stop, rewind, etc) stays on the smartphone since the HiFi system is not suitable for displaying the movie.

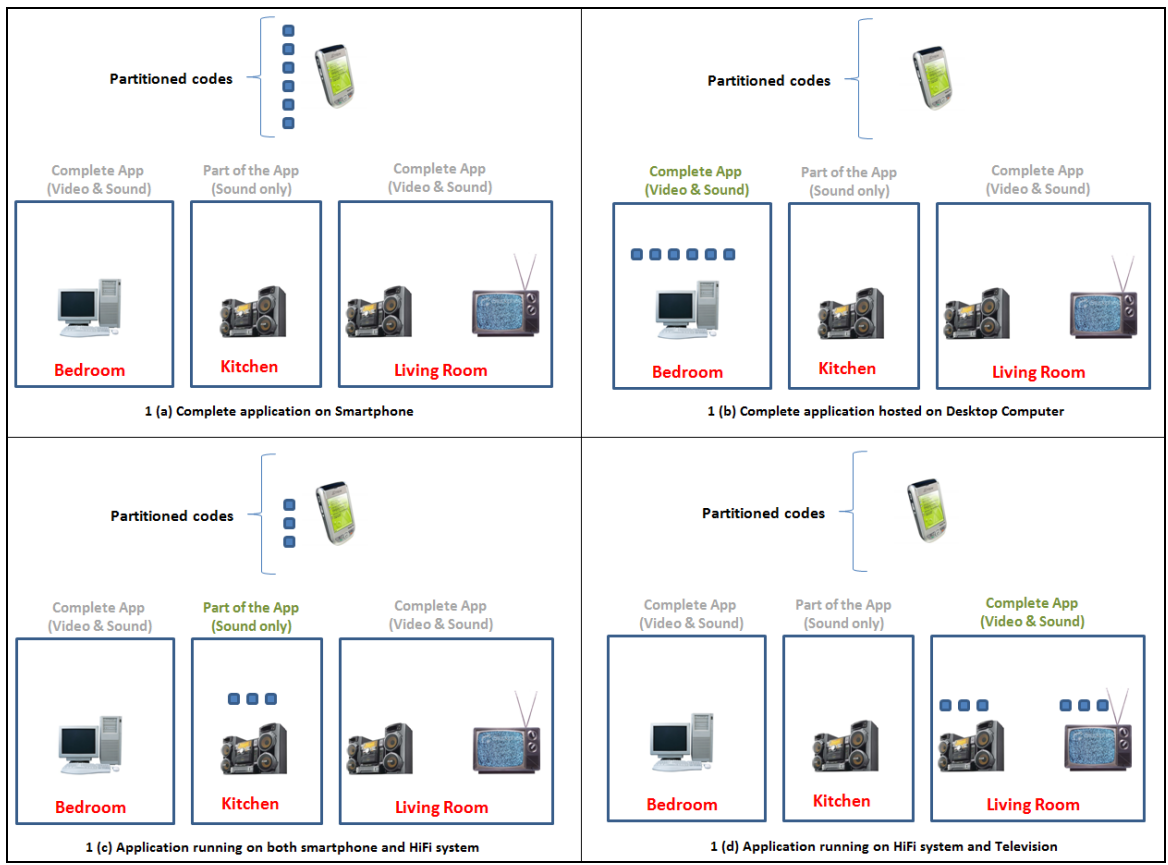

Figure 1. Video player application running in a pervasive environment

Finally, the user decides to go to the living room, situation 1(d) and the same process again. The application retract back on his smartphone and once in his living room, a HiFi system and television are detected, the video player decompose / partition itself and move part of the partitioned software to the HiFi system and part of it to the television set.

In this simple example that demonstrates the interaction of the user, software and device in a pervasive environment, we identify several important and interesting issues. First there is a sort of autonomous behavior, w.r.t. the user's preferences that the application possesses. It is no longer required unlike traditional environment or software to stop the application on one device and start it on another; instead maximum use of the resources in the surrounding environment is being made.

Having a copy of each part of the partitioned application on the smartphone ensures continued availability of the application in case of network disconnection due to the user mobility. Thus the retraction property is guaranteed.

Though it is not mentioned above, devices/nodes' privileges is a very important design principle to consider. In real-life pervasive environment, unlike the scenario of the music player where there is only one user and one application, we are likely to be exposed to multiple users and multiple applications running concurrently in the environment, using and releasing resources at any time. For example, in situation 1(d), if a second user enters the living room and has higher priority than the first user over the television and want to use that resource, then the parts of the application that is running on the television needs to retract back to the smartphone and continue playing the movie. There are lots of other scenarios where the second user can allow the first user to continue enjoying the resource (i.e., television), and so on, and different configurations can be applied. However, the basic idea is to understand the importance and incorporation of the device/node privileges principle in the framework.

IV. PARTITIONING AND RETRACTION FRAMEWORK

Figure 2 below shows how the BubbleCodes framework works. In short, a context-aware application running on a smartphone is partitioned and a copy of each partitioned block of codes referred as a *Bubble* is moved to available nodes in the environment. Each available node, which can be a television, desktop computer, HiFi system, etc, can host one or more Bubbles.

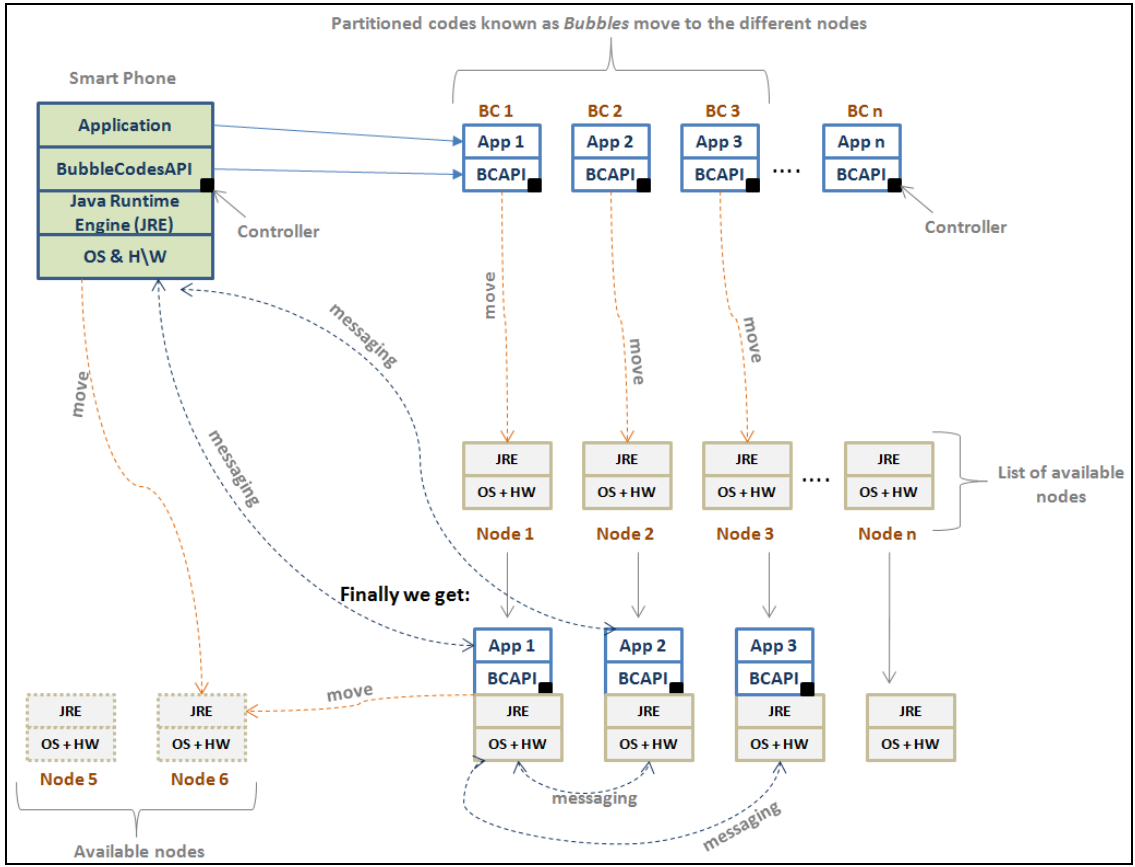

Figure 2. BubbleCodes Framework

The idea of moving a copy of the Bubbles to the nodes ensures network resilience and full retraction. Since whenever a node hosting a Bubble is out of network or simply fail due to any hardware problem (e.g., low power), we always have a copy of the Bubble on the source device.

*A. Controller*

The Controller is responsible for managing the different operations from partitioning to retraction. At design time the designer or application developer will have to define a Controller for an application. The Controller on the source device is referred as $C_0$ and while other Controllers are referred to as $C_1$, $C_2$, etc. $C_0$ will have to continuously monitor the availability of the nodes hosting the Bubbles. In case of node unavailability due to privileges as we discussed above in Section III, $C_0$ will inform the Bubble to change its hosting node. A specific node can also be recommended. $C_1$, $C_2$, etc, have to continuously update $C_0$ about their status during runtime. The Controllers identify and remember each node and Bubble by their unique name. This facilitates management of nodes assignment, nodes switching, etc, in multi-user and multi-application environments.

*B. Partitioning Process*

Once launch the application on the source device will first attempt to detect available node s in its surrounding environment; then check for permission and verify the "qualifications" of the nodes. Qualifications of nodes refer to the capabilities of the latter, for e.g., some have display capability, sound output, input capability and so on. A copy of the Bubbles is then moved to the selected nodes along with a *BubblesMap*.

*C. Bubbles Mapping*

Once the application has been partitioned, and the nodes to host each Bubble has been selected, a *BubblesMap* is created that list all Bubbles and their corresponding nodes and all available but unused nodes; and is sent to all nodes hosting a Bubble in that list. This has three important advantages:

- It allows $C_0$ to know exactly where a Bubble is located at any point in time.

- In case of node unavailability due to privileges as we discussed above, the node can choose any available node in the *BubblesMap* list, automatically move to that node, update the *BubblesMap* and send an updated copy of the latter to the $C_0$; it will be sent distributed to all nodes concerned.

- Also, if a node fails, $C_0$ can choose another node to send a copy of the Bubble.

Whenever there is an update in the *BubblesMap*, a copy of the latter is sent to all Bubbles concerned and to $C_0$ if the update was not generated by it.

*D. Bubbles*

As specified earlier, Bubbles are the different blocks of codes of the application. A Bubble can typically be divided into two distinct parts:

- First is the part of the application's codes,
- And second, is the part of the BubbleCodes framework.

Operations performed by Bubbles are as follows:

- *Listen for context variable change*: Since the application is a context-aware one, Bubbles have to listen to any change in their respective context variable. This can be achieved in three ways, either $C_0$ update the Bubble about the context change or the Bubble detect it by itself; or finally a hybrid approach combining the above two methods.
- *Adapt to context change*: The Bubble will have to adapt autonomously and automatically its operations to any change in context.
- *Switch nodes*: In case of privileges situation, the Bubble has to find another node and switch itself to it. The node that requires minimum communication time is given higher priority.
- *Communicate with colleagues*: A Bubble can also communicate as and when required with any other Bubble present in its *BubblesMap* to exchange information.
- *Self-Destruction*: Upon completion of its task a Bubble informs $C_0$ or vice-versa and in case of node switching, a Bubble needs to destruct itself to deallocate memory space and stop any computation activity.

*E. Retraction Process*

The retraction process can be executed in three ways. First, $C_0$ inform a Bubble's $C_1$ that it needs to stop running and to self-destroy. Second, a Bubble can inform $C_0$ that it is stopping execution, for e.g., due to end of task/job. Third, $C_0$ stops detecting a Bubble, for e.g., due to network disconnection because of users mobility. In all above cases, immediately after that, $C_0$ will now only communicate with the Bubble at the source device and there will be an update in the *BubblesMap*.

*F. Messaging*

*Messaging* is the exchange of information between the components. There as several types of message that are exchanged between the $C_0$ and $C_1$, $C_2$, and so on and Bubbles among Bubbles:

- Sending or receive *BubblesMap*: Both $C_0$ and Bubbles can send or receive *BubblesMap*.
- Send or receive context information: Both $C_0$ and Bubbles can send or receive context information to adapt the application's behavior.
- Send or receive application's data: Both $C_0$ and Bubbles can send or receive application's data.
- Request Permission: Before the source device moves a Bubble to a node or a node moves a Bubble to another node, permission must be requested to the new node.
- Request Node Qualifications: Message is sent to a node to check if the latter is qualified, that is, if it hold the minimum requirements of a Bubble in terms of hardware capabilities.
- Bubble task completion: For example, once its task completed, a Bubble will inform $C_0$.

V. DISCUSSIONS

Developing a fully functional or full-fledge framework that would ensure complete reliability of the partitioning and retraction of the Bubbles is not a straight forward process and a lengthy one. In the previous section, we integrate the different underlying principles we identified so far. As we built applications on top of BubbleCodes, we will try to learn new design principles or improve existing ones to eventually improve the framework itself.

Message passing is an important feature of BubbleCodes. All messages exchanged by the different components will be implemented in XML and a parser at each end will process them. Using XML helps to reduce the size of the message exchange and yet transport valuable information.

The choice of the language is normally independent of the framework. However, considering the popularity of the Java programming language and the release of the Android OS [20] on smartphones by many major key manufacturers and the billions of java enabled mobile phones, these factors motivate us to proceed ahead with the framework implementation in Java.

Concerning the selection of the communication technology, we have the choice between Bluetooth and WiFi. The choice for our first implementation will most probably be in favor of WiFi but again the framework is independent of communication technology used.

As mentioned earlier in Section IV, the designer or application developer will have to define the Bubbles. In that we mean that he/she needs to specify which part of the codes can be partitioned and also describe the minimum requirements a node needs to host that particular Bubble. This technique has been used by several other works like in J-Orchestra [18] the authors used the keyword *mobile* to specify classes that move to other nodes and in Doorastha [16] the authors used *copyable* (objects will be transferred by-copy) or *migratable* (objects will be migrated to a node).

## VI. CONCLUSIONS AND FUTURE WORKS

Even if several researches have been around for years in the area of software partitioning, most of them do not have the profile to be considered for usage in mobile pervasive environments. To our knowledge, currently the work in [13][14] is the only step forward in this direction. However, we identified several important issues not taken into consideration to approach partitioning and retraction. In this paper we propose BubbleCodes which is an application partitioning and retraction framework for pervasive environments. We first describe a scenario to explain the benefits of incorporating the properties of partitioning and retraction in next-generation context-aware applications. We then proceed with a detailed explanation about how the BubbleCodes framework works. Next we intend to proceed with the architectural design and implementation of the framework to finally built test applications on top of it. Likewise, we will be able to further improve and fine-tune BubbleCodes so that application developers can easily built context-aware applications since the BubbleCodes framework will manage all the low-level complexities that currently application developers have to deal with.

## VII. ACKNOWLEDGEMENT

## REFERENCES

[1] R. Want, A. Hopper, V. Falcao and J. Gibbons, "The Active Badge Location System", ACM Transactions on Office Information Systems (TOIS), Vol. 10. No. 1, January 1992.

[2] M. Weiser, "The Computer for the Twenty-First Century," Scientific American, September 1991.

[3] Google Scholar, http://www.scholar.google.com, retrieve on 08th September 2010.

[4] B. N. Schilit, N. Adams and R. Want, "Context-Aware Computing Applications" IEEE Workshop on Mobile Computing Systems and Applications, December 1994.

[5] R. Ait Yaiz, F. Selgert and F. den Hartog, "On the definition and relevance of context-awareness in Personal Networks", in proceedings of the Third Annual International Conference on Mobile and Ubiquitous Systems: Networking & Services, July 2006.

[6] E. J. Y. Wei and A. T. S. Chan, "Towards Context-Awareness in Ubiquitous Computing", in proceedings of the 2007 International Conference on Embedded and Ubiquitous Computing, 2007.

[7] A. Zimmermann, A. Lorenz and R. Oppermann, "An Operational Definition of Context", Lecture Notes in Computer Science, Volume 4635/2007, 2007.

[8] A. K. Dey, "Understanding and Using Context", Personal and Ubiquitous Computing, Volume 5, Issue 1, Springer-Verlag, February 2001.

[9] T. Zhang, "An Architecture for Building Customizable Context-Aware Applications by End-Users", in proceedings on the Second International Conference, PERVASIVE 2004, 2004.

[10] A. Wood, J. Stankovic, G. Virone, L. Selavo, Z. He, Q. Cao, T. Doan, Y. Wu, L. Fang and R. Stoleru, "Context-Aware Wireless Sensor Networks for Assisted-Living and Residential Monitoring", in proceedings of the IEEE Network, 2008.

[11] N. S. Park, K. W. Lee and H. Kim, "A Middleware for Supporting Context-Aware Services in Mobile and Ubiquitous Environment", in proceedings of the International Conference on Mobile Business (ICMB'05), July 2005.

[12] G. Chen and D. Kotz, "Context Aggregation and Dissemination", in proceedings of the 4th IEEE Workshop on Mobile Computing Systems and Applications, June 2002.

[13] J. Vallejos, E. Gonzalez Boix, E. Bainomugisha, P. Costanza, W. De Meuter, and É. Tanter, "Towards Resilient Partitioning of Pervasive Computing Services", in proceedings of the 3rd ACM Workshop on Software Engineering for Pervasive Services, July 2008.

[14] E. Bainomugisha, J. Vallejos, É. Tanter, E. Gonzalez Boix, P. Costanza, W. De Meuter, and T. D'Hondt, "Resilient Actors: A Runtime Partitioning Model for Pervasive Computing Services", in proceedings of the 2009 International Conference on Pervasive Services, ICPS'09, July 2009.

[15] M. Philippsen and M. Zenger, "JavaParty – transparent remote objects in Java", Concurrency: Practice and Experience, 9(11):1225-1242, November 1997.

[16] M. Dahm, "Doorastha: a step towards distribution transparency", in Net. ObjectDays, 2000.

[17] M. Tatsubori, T. Sasaki, S. Chiba and K. Itano, "A Bytecode Translator for Distributed Execution of "Legacy" Java Software", in proceedings of the 15th European Conference on Object Oriented Programming, (ECOOP 2001), June 2001.

[18] E. Tilevich and Y. Smaragdakis, "J-Orchestra: Enhancing Java Programs with Distribution Capabilities", in proceedings of the ACM Transactions of Software Engineering and Methodology, Vol. 19, No. 1, August 2009.

[19] I. Satoh, "Dynamic deployment of pervasive services", in proceedings of the International Conference Pervasive Services, ICPS'05, 2005.

[20] Android OS coverage on Wikipedia (accessed on 09th October 2010): http://en.wikipedia.org/wiki/Android_(operating_system)